# Method for backtracking the layer thermal conductivities of multilayer thin film structure using coupled Newton Raphson approach and 3$\omega$ method

*Aigbe E. Awenlimobor, Jiajun Xu*
*Dept. of Mechanical Engineering, University of the District of Columbia, D.C , 20008, USA*

## Abstract

The thermal conductivity of thin films is commonly estimated using the 3-omega experimental method. When calibrating the test setup, it is customary to use a specimen with a known thermal conductivity for validation. However, when determining the thermal conductivity of samples with unknown values, numerical approximations can provide a means to validate experimental results and ensure the integrity of the setup. A simple analytical or finite element analysis (FEA) method can be used to achieve this. For multilayer systems of unknown layer thermal conductivities, the 3-omega experimental setup only provides information about the overall bulk thermal conductivity of the system. To obtain the individual layer thermal conductivities, a combined experimental and numerical approach can be used. This article presents a novel method for backtracking the layer thermal conductivities of a multilayer thin film structure using a coupled 3-omega experimental and Newton-Raphson numerical approach. The method is validated using high-fidelity data obtained from literature.

## Theory

A typical sample preparation for the 3-omega setup is shown in Figure 1 (a) which comprise of a nanostrip heater element, centered on a thin film with unknown $\kappa_f$ which is padded onto a substrate.

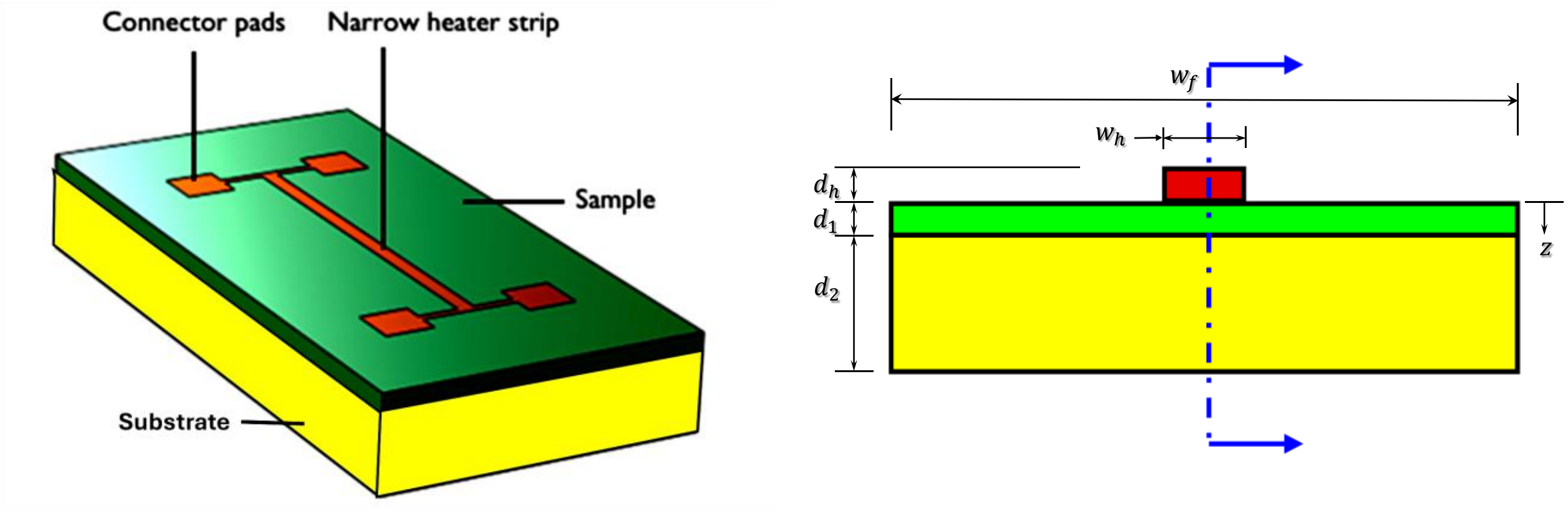

(a) (b)

Figure 1: (a) Schematic showing typical 3$\omega$ method sample preparation setup (b) two-dimensional cross section of the experimental setup.

The nano strip is thermally excited with a sinusoidal alternating current (ac) of the form $I = I_0 \cos(\omega t)$, heat is generated via resistance of the nanostrip elements and the power dissipated is given as [3]

$$P = I_0^2 R_0 \cos^2(\omega t) = P_0[1 + \cos(2\omega t)] \qquad 1$$

where

where $P_0 = P_{rms} = I_0^2 R_0/2$ is the root mean square (rms) power component dissipated per ac cycle, $R_0$ is the nanostrip heater resistance at nominal temperature $T_0$, given as $R_0 = \rho_{th}\, l_h/A_\perp$, where $A_\perp$ is the in-plane area of the nanostrip heater given as $A_{\perp} = w_h d_h$, $w_h$ is the heater width, $d_h$ is the heater height, $l_h$ is the length of the nanostrip heater and $\rho_{th}$ is its electrical resistivity. The angular modulation frequency $\omega$ is given as $\omega = 2\pi f$, where $f$ is the excitation frequency in $Hz$. For small fluctuations in temperature ($\Delta T$), the nanostrip heater resistance, $R$ can be derated according to the equation [1,2,3,4]

$$R = R_0(1 + \beta\Delta T) \tag{2}$$

where $\beta$ is the nanostrip heater resistance temperature derating factor. The temperature oscillation across the thin-film specimen reckoned from the heater/film interface at $z = 0$ (cf. Figure 1 (b)) is given as

$$\Delta T(t,z) = \Delta T_0 e^{i2\omega t} e^{-qz} \tag{3}$$

Where the oscillating heater temperature $\Delta T_0$ is given as $\Delta T_0 = \left|\overrightarrow{\Delta T}\right|_{ac} e^{-i\pi/4}$ [3], $\left|\overrightarrow{\Delta T}\right|_{ac} = \left(P_0/\kappa\, A_{||}\right)(2\omega/\alpha)^{-1/2}$ is the peak amplitude of the ac temperature oscillations, $A_{||}$is the cross plane area of the nanostrip heater given as $A_{||} = l_h w_h$, $q$ is the wavenumber given as $q = \left({}^{2\omega}/_{\alpha}\right)^{1/2} e^{i\pi/4}$, $\alpha$ is the film's thermal diffusivity ($\alpha = \kappa/\rho C_p$). The derated heater resistance as a function of time at $z = 0$ is given as [3]

$$R(t) = R_0\left[1 + \beta\left\{\Delta T_{dc} + \left|\overrightarrow{\Delta T}\right|_{ac}\cos(2\omega t + \pi/4)\right\}\right] \tag{4}$$

where $|\Delta T|_{dc}$ is the temperature rise due to the component of the direct current (dc). The voltage across the nanostrip heater element is given as [3]

$$V(t) = V_0\left[(1 + \beta\Delta T_{dc})\cos(\omega t) + \frac{1}{2}\beta\left|\overrightarrow{\Delta T}\right|_{ac}\cos\left(\omega t + \frac{\pi}{4}\right) + \frac{1}{2}\beta\left|\overrightarrow{\Delta T}\right|_{ac}\cos\left(3\omega t + \frac{\pi}{4}\right)\right] \tag{5}$$

where $V_0 = I_0 R_0$. The amplitude of the third summand which has the $3\omega$ component represents the $V_{3\omega}$ voltage, i.e.

$$V_{3\omega} = \frac{1}{2}V_0\beta\left|\overrightarrow{\Delta T}\right| \tag{6}$$

Given a thermal penetration depth $d_{th}$ into the film sample, the required thermal wave oscillation frequency is given as $\omega_{th} = 2\pi\,\alpha/d_{th}^2$. Formulations of the temperature oscillation for one-dimensional line nanostrip heater at the center of an infinite cylindrical specimen and at the mid-surface of an infinite half - cylindrical specimen are provided in the Appendix [3], and likewise for a finite width heater at the mid-surface of an infinite half - cylindrical specimen padded onto a substrate. For multilayer system, with the substrate as the n$^{th}$ layer, the average temperature oscillation amplitude of the nanostrip heater neglecting effects of the interface thermal resistances and thermal storage capacity of the heater element can be given by the complex integral [1,2]

$$\overrightarrow{\Delta T}_{ac} = -\frac{P_{rms}}{\pi l_h \kappa_{||_1}}\int_0^{\infty}\frac{1}{A_1 B_1}\frac{\sin^2(w_h\lambda/2)}{w_h^2\lambda^2/4}\,d\lambda \tag{7}$$

where

$$A_{i-1} = \frac{A_i\dfrac{\kappa_{||_i}B_i}{\kappa_{||_{i-1}}B_{i-1}} - \tanh(\varphi_{i-1})}{1 - A_i\dfrac{\kappa_{||_i}B_i}{\kappa_{||_{i-1}}B_{i-1}}\tanh(\varphi_{i-1})}, \qquad B_i = \left(\frac{\kappa_{\perp i}}{\kappa_{||_i}}\lambda^2 + \frac{j2\omega}{\alpha_{||_i}}\right)^{1/2}, \qquad \varphi_i = B_i d_i, \tag{8}$$

$$i = 2 \ldots n$$

index $i = 1$ corresponds to the top film layer and index $n$ corresponds to the substrate layer. For boundary conditions at the substrates bottom surface determines the value $A_n$ and is summarized thus:

$$A_n = \begin{cases} -1 & z_n \to \infty \\ -\tanh(\varphi_n) & \left.\frac{\partial T}{\partial z}\right|_n = 0 \\ -[\tanh(\varphi_n)]^{-1} & T_n = const. \end{cases} \quad 9$$

In equations 8 above, $d_i$ is the layer thickness, $\kappa_i$ is the layer 's thermal conductivity ($\perp$ corresponds to the cross-plane direction, i.e. direction perpendicular to the film/substrate interface, and $||$ corresponds to the in-plane direction), $\alpha_i$ is the layer's thermal diffusivity. If the average temperature oscillation amplitude at the nanostrip heater/film interface $\overline{|\Delta T|}$ or the $V_{3\omega}$ is known from experiments for different excitation frequencies $\omega$, then the average thermal conductivity $\kappa$ of the specimen can be calculated using the expression given as

$$-\frac{2\pi l_h}{P_{rms}}\kappa_{||} = \frac{\partial(\ln \omega)}{\partial\left(\left|\overrightarrow{\Delta T}\right|\right)} \approx \frac{V_0\beta}{2}\frac{\partial(\ln \omega)}{\partial(V_{3\omega})} \quad 10$$

If the thermal conductivities of the specimens' different strata are known a priori during a calibration exercise, then $\kappa_{||}$ can be validated with that computed from the average $\bar{\kappa}_{||} = \sum \kappa_{||_i} d_i / \sum d_i$.

## Numerical Simulation

As an alternative method to analytical approximations, the numerical FE method can be used to estimate the average temperature oscillation amplitude $\left|\overrightarrow{\Delta T}\right|$ and the $V_{3\omega}$ to validate experimental results. A typical half-symmetry model of the $3\omega$ method specimen setup is shown in Figure 2 below. An infinite domain boundary condition is assumed at the bottom of the substrate and the right end of the specimen. For the preliminary exercise, convective heat loss at the specimen's top surface and the thermal resistance at the layer interfaces are neglected.

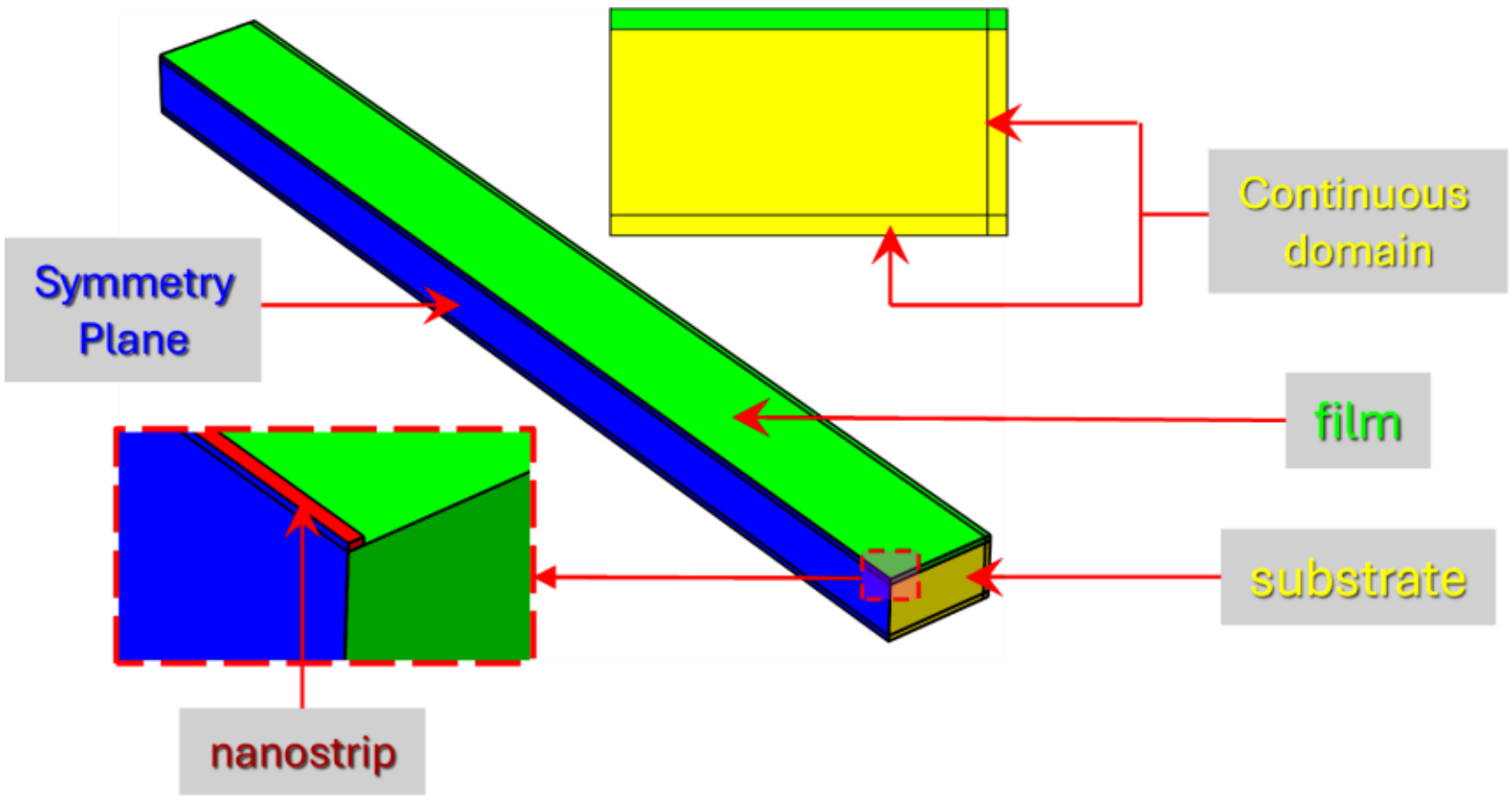

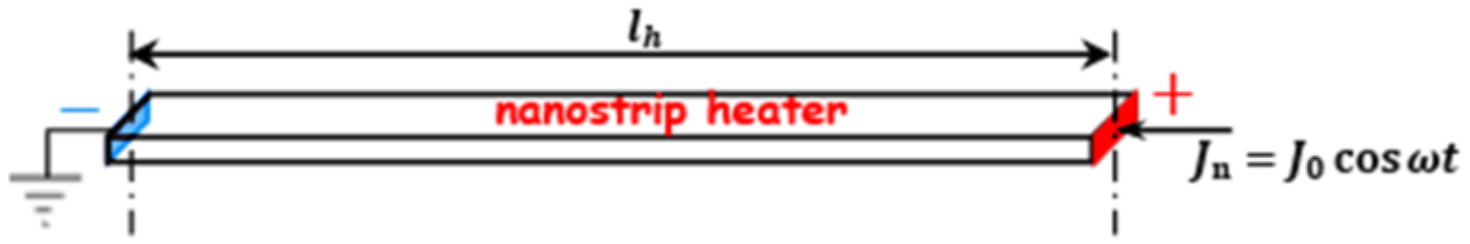

Figure 2: Half symmetry model of the 3$\omega$ method specimen configuration showing relevant boundary conditions

The length of the specimen used is given as $l_h = 1\ mm$ while all other relevant dimensions of the specimen configuration are given in Table 1 below.

Table 1: Geometry of the specimen layup

| | **Width $w$ [$\mu m$]** | **Thickness $d$ [$\mu m$]** |
|---|---|---|
| nanostrip heater | 0.4 | 0.1 |
| Thin Film | 100 | 5.0 |
| Substrate | 100 | 50. |

The material properties of the nanostrip heater element, the thin film and the substrate used for the current study were obtained from Ref. [1] and presented in Table 2 below.

Table 2: Material properties of the specimen layup [1]

| | **Material** | **$\kappa$ [W/m − K]** | **$\rho$ [kg/$m^3$]** | **$C_p$ [J/kg − K]** |
|---|---|---|---|---|
| nanostrip heater | Au | 318 | 19300 | 129 |
| Thin Film | AIN | 285 | 3260 | 740 |
| Substrate | Si | 142 | 2330 | 710 |

The nanostrip resistance temperature derating factor is given as $\beta = 0.0038$ and the reference resistivity of the nanostrip heater is given as $\rho_{th} = 1.58 \times 10^{-8} \Omega \cdot m$. The nominal heating resistance of the nano strip is calculated from $R_0 = \rho_{th} l_h / A_{\perp}$. The FEA simulation is carried out using COMSOL Multiphysics 6.3 software, (COMSOL Inc, Burlington, MA, USA). An inward normal current flux is prescribed at one in-plane face of the nanostrip with a normal flux density given as $J_n = J_0 \cos \omega t$ where $J_0 = I_0 / A_{\perp}$ where $I_0 = 60 mA$, while the opposite end face is grounded and all other nanostrip free surfaces are electrically insulated. Heat source is defined in the entire nanostrip domain at a heat rate of $\dot{Q}_s = P_0/2 = I_0^2 R_0 / 4$. In advanced studies, convective heat flux with a heat transfer coefficient $h_t$ [W/m$^2$K] will be specified and thermal resistance at the interfaces of contacting surface will be specified using the 'thin structures'-'thermal contact' boundary physics. A temperature boundary condition at ambient temperature value is prescribed at the bottom surface of the substrate while all other exposed faces are thermally insulated. Two case studies are considered for the simulation, Case A. – substrate having the same material properties as the thin film, and Case B. each layer has its unique material properties. A parametric transient study is carried out at different current excitation frequencies ranging from 1 $[Hz] \leq \log_{10} f \leq 5$ $[Hz]$ in steps of 0.5 $[Hz]$ and subsequently frequency study is carried out using the forward fast fourier transform (FFT) solver to transform solutions from time domain to frequency domain with a maximum output frequency up to at least three (3) times the nominal excitation frequencies to obtain the $V_{3\omega}$ solutions. Simple linear regression analysis is carried out using the $V_{3\omega}$ output data at the various $\omega$, and the gradient of the fitted line $m = \partial(V_{3\omega})/(\ln \omega)$ is obtained from the regression coefficients. The specimen's average thermal conductivity is thus computed from the slope according to the equation 10 above.

**Obtaining Individual Layer Thermal Conductivities**

A potential useful application of equations 7-9 is the ability to backtrack the layer thermal conductivities given the excitation frequencies, $\omega$ and associated temperature oscillation amplitudes $|\overrightarrow{\Delta T}|$, or $V_{3\omega}$ voltages can be determined from experiments. To achieve this, a simple numerical method such as the well-known Newton-Raphson (NR) algorithm could be used. The NR iterative approach to determine the $\kappa_{||}$ values for a multilayer system can be written as

$$\kappa_{\|_i}^{+} = \kappa_{\|_i}^{-} + J_{ij}^{-\,-1} R_j^{-} \tag{11}$$

where the superscripts $(+)$ and $(-)$ denotes the current and previous iterations, $R_j^-$ is the NR residual vector defined in this case as $R_j^- = |\overrightarrow{\Delta T}|_j^- - |\overrightarrow{\Delta T}|_j^{exp}$ corresponding to the $j^{th}$ frequency, and $J_{ij}^-$ is the Jacobian given as

$$\begin{aligned} J_{ij}^- &= \frac{\partial R_j^-}{\partial \kappa_{\|_i}^-} = \frac{\partial |\overrightarrow{\Delta T}|_j^-}{\partial \kappa_{\|_i}^-} = \frac{1}{|\overrightarrow{\Delta T}|} \overrightarrow{\Delta T}_j^- \cdot \frac{\partial \overrightarrow{\Delta T}_j^-}{\partial \kappa_{\|_i}^-} \\ &= \frac{1}{|\overrightarrow{\Delta T}|}\left[ Re\left(\overrightarrow{\Delta T}_j^-\right) Re\left(\frac{\partial \overrightarrow{\Delta T}_j^-}{\partial \kappa_{\|_i}^-}\right) + Im\left(\overrightarrow{\Delta T}_j^-\right) Im\left(\frac{\partial \overrightarrow{\Delta T}_j^-}{\partial \kappa_{\|_i}^-}\right)\right] \end{aligned} \tag{12}$$

The iteration is terminated once a preset error tolerance is achieved, which in this case is defined as the Euclidean norm of the residual, i.e. $err = \|R_j^-\|$. To obtain the derivatives of $\overline{|\Delta T|}_i$ with respect to the layer thermal conductivities $\kappa_{\|_j}$, it is convenient to recast the equation 7 to the form given as

$$\overrightarrow{\Delta T}_i = -\frac{P_{rms}}{\pi l_h \kappa_{\|_1}} \int_0^\infty f_i\left(\kappa_{\|_j}, \lambda\right) d\lambda, \qquad f_i\left(\kappa_{\|_j}, \lambda\right) = \frac{1}{A_1 B_1} \frac{\sin^2(w_h \lambda/2)}{w_h^2 \lambda^2/4} \tag{13}$$

such that

$$\begin{aligned} &A_i = \tanh(\psi_i - \varphi_i), \qquad B_i = \left(\lambda^2 \kappa_{R_i} + \frac{j2\omega}{\alpha_{\|_i}}\right)^{1/2} \\ &\psi_i = \operatorname{atanh}\left(A_{i+1}\frac{\kappa_{\|_{i+1}} B_{i+1}}{\kappa_{\|_i} B_i}\right), \qquad \varphi_i = B_i d_i, \qquad \kappa_{R_i} = \frac{\kappa_{\perp_i}}{\kappa_{\|_i}}, \qquad \alpha_{\|_i} = \frac{\kappa_{\|_i}}{\rho_i C_{p_i}} \\ &i = 1 \ldots n-1 \end{aligned} \tag{14}$$

where all quantities retain their usual meaning. To obtain the gradient of the integral function $\overline{|\Delta T|}_i$ w.r.t. $\kappa_{\|_j}$The Leibnitz integral rule is utilized in combination with the product rule, i.e.

$$\begin{aligned} \frac{d}{d\kappa_{\|_j}} \overrightarrow{\Delta T}_i &= -\delta_{1j} \frac{1}{\kappa_{\|_1}} \overrightarrow{\Delta T}_i \\ &- \frac{P_{rms}}{\pi l_h \kappa_{\|_1}} \left[ \int_0^\infty \frac{\partial}{\partial \kappa_{\|_j}} f_i\left(\kappa_{\|_k}, \lambda\right) d\lambda + f_i\left(\kappa_{\|_k}, \infty\right) \frac{\partial \infty}{\partial \kappa_{\|_j}} - f_i\left(\kappa_{\|_k}, 0\right) \frac{\partial 0}{\partial \kappa_{\|_j}} \right] \end{aligned} \tag{15}$$

Fortunately, the integral limits are constants and $\partial\infty/\partial\kappa_{\|_j} = \partial 0/\partial \kappa_{\|_j} = 0$. The derivative of the function $f_i\left(\kappa_{\|_k}, \lambda\right)$ can be expanded to give

$$\frac{\partial}{\partial \kappa_{\|_j}} f_i\left(\kappa_{\|_k}, \lambda\right) = -f_i\left(\kappa_{\|_k}, \lambda\right)\left[\frac{1}{A_1}\frac{\partial A_1}{\partial \kappa_{\|_j}} + \frac{1}{B_1}\frac{\partial B_1}{\partial \kappa_{\|_j}}\right] \qquad 16$$

where

$$\frac{\partial A_i}{\partial \kappa_{\|_j}} = \left[1 - A_i^2\right]\left[\frac{\partial \psi_i}{\partial \kappa_{\|_j}} - \frac{\partial \varphi_i}{\partial \kappa_{\|_j}}\right] \qquad 17$$

$$\frac{\partial B_i}{\partial \kappa_{\|_j}} = -\frac{1}{2}\delta_{ij}\begin{cases} \dfrac{B_j}{\kappa_{\|_j}}, & \kappa_{R_j} = \kappa_{R_j}\left(\kappa_{\|_j}\right) \\ \dfrac{1}{B_j \kappa_{\|_j}}\left[\dfrac{j2\omega}{\alpha_{\|_j}}\right], & \kappa_{R_j} = const \end{cases} \qquad 18$$

The derivative of the term $\partial\psi_i/\partial\kappa_{\|_j}$ is recursive in nature since it depends on the derivatives of the downstream layers, i.e. $\partial A_i/\partial\kappa_{\|_j}$ and is given as

$$\frac{\partial \psi_i}{\partial \kappa_{\|_j}} = \frac{\sinh(2\psi_i)}{2}\begin{cases} -\left[\dfrac{1}{\kappa_{\|_j}} + \dfrac{1}{B_j}\dfrac{\partial B_j}{\partial \kappa_{\|_j}}\right] & j = i \\ \left[\dfrac{1}{\kappa_{\|_j}} + \dfrac{1}{A_j}\dfrac{\partial A_j}{\partial \kappa_{\|_j}} + \dfrac{1}{B_j}\dfrac{\partial B_j}{\partial \kappa_{\|_j}}\right] & j = i + 1 \\ \dfrac{1}{A_{i+1}}\dfrac{\partial A_{i+1}}{\partial \kappa_{\|_j}} & i + 1 < j < n - 1 \end{cases} \qquad 19$$

The derivative $\partial\varphi_i/\partial\kappa_{\|_j}$ is simply given as

$$\frac{\partial \varphi_i}{\partial \kappa_{\|_j}} = \frac{\partial B_i}{\partial \kappa_{\|_j}} d_j \qquad 20$$

The last layer derivative $\partial A_n/\partial\kappa_{\|_n}$ where the recursion terminates depends on base surface boundary condition and is given as

$$\frac{\partial A_n}{\partial \kappa_{\|_n}} = \begin{cases} 0 & z_n \to \infty \\ -\operatorname{sech}^2(\varphi_n)\dfrac{\partial \varphi_n}{\partial \kappa_{\|_n}} & \left.\dfrac{\partial T}{\partial z}\right|_n = 0 \\ \operatorname{csch}^2(\varphi_n)\dfrac{\partial \varphi_n}{\partial \kappa_{\|_n}} & T_n = const. \end{cases} \qquad 21$$

For $n$ layers system, we require $n$ number of experiments, i.e. $n$ frequencies ($\omega$) and associated $\overline{|\Delta T|}$, or $V_{3\omega}$ values. To demonstrate this, a three-layer system is utilized, with a nanofilm layer having a thickness of $10.0\mu m$ and width of $100\mu m$ inserted between the AlN film and Si substrate and having properties given in Table *3* below [4].

Table 3: Material properties of the additional film layer [4]

| | **Material** | $\boldsymbol{\kappa}$ **[W/m − K]** | $\boldsymbol{\rho}$ **[kg/$m^3$]** | $\boldsymbol{C_p}$ **[J/kg − K]** |
|---|---|---|---|---|
| nanofilm | Unknown | 120 | 3100 | 820 |

As with any other iterative root finding method, the initial guess value largely determines the ability of the method to yield accurate and physical real roots. A good initial guess would be the average conductivity values determined from the $V_{3\omega}$ experiment.

## Preliminary Results and Discussion

### Obtaining the Average System Thermal Conductivities

The results of studies including the average thermal oscillation transient profiles, $\Delta T$,linear regression plots of the $V_{3\omega}$ voltages and average temperature oscillation amplitudes $|\overrightarrow{\Delta T}|$ for the different excitation frequencies, $\omega$, and plots comparing the analytical and numerical regression output profiles are presented in the subsections following for both cases i.e. case A – considering same material properties for both substrate and thin film layers & case B – considering different properties for both layers.

### 1. Case A – *Same material properties for both substrate and thin film layers*

Figure 3 (a) shows a typical temperature distribution at an excitation frequency of $f = 10Hz$ and at times $t = 0.4s$. As expected, the temperature is relatively higher near the nanostrip heating element and diffuses as we move farther away from heater to the insulated surfaces. Likewise, the peak amplitude of the temperature oscillations is seen to increase monotonically and steadily with the excitation frequency (cf. Figure 3 (b)).

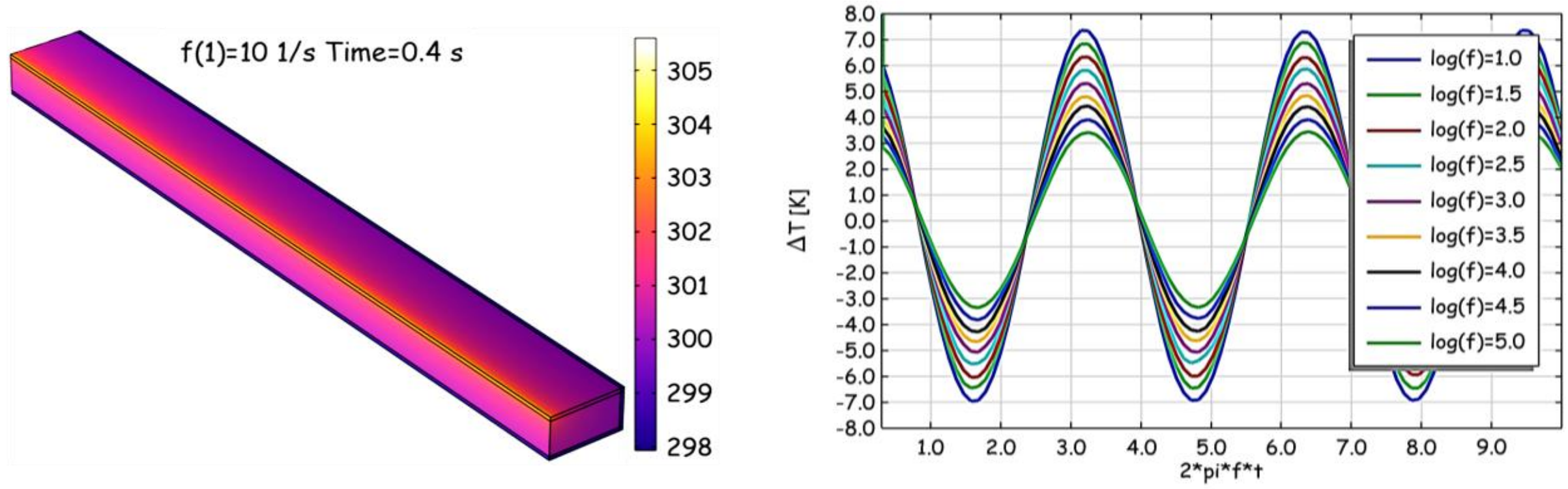

(a) (b)

Figure 3: Case A- (a) Typical temperature distribution of the sample preparation at a frequency of $f = 10Hz$ and at time $t = 0.4s$, (b) temperature oscillations transient profiles for the different excitation frequencies with the times axis normalized with the frequency (i.e. $\omega t$).

Figure 4 (a) & (b) compare numerical and analytical results of the linear regression analysis for the $V_{3\omega}$ voltages and the peak oscillation temperature amplitudes $\overline{|\Delta T|}$ obtained from the frequency based parametric studies. For both outputs, it is observed that the analytical solutions underpredict the responses compared to the numerical solutions. This may be due to the fact that the analytical equations are derived from a one-dimensional (1D) space assumption unlike the full three three-dimensional (3D) numerical simulations.

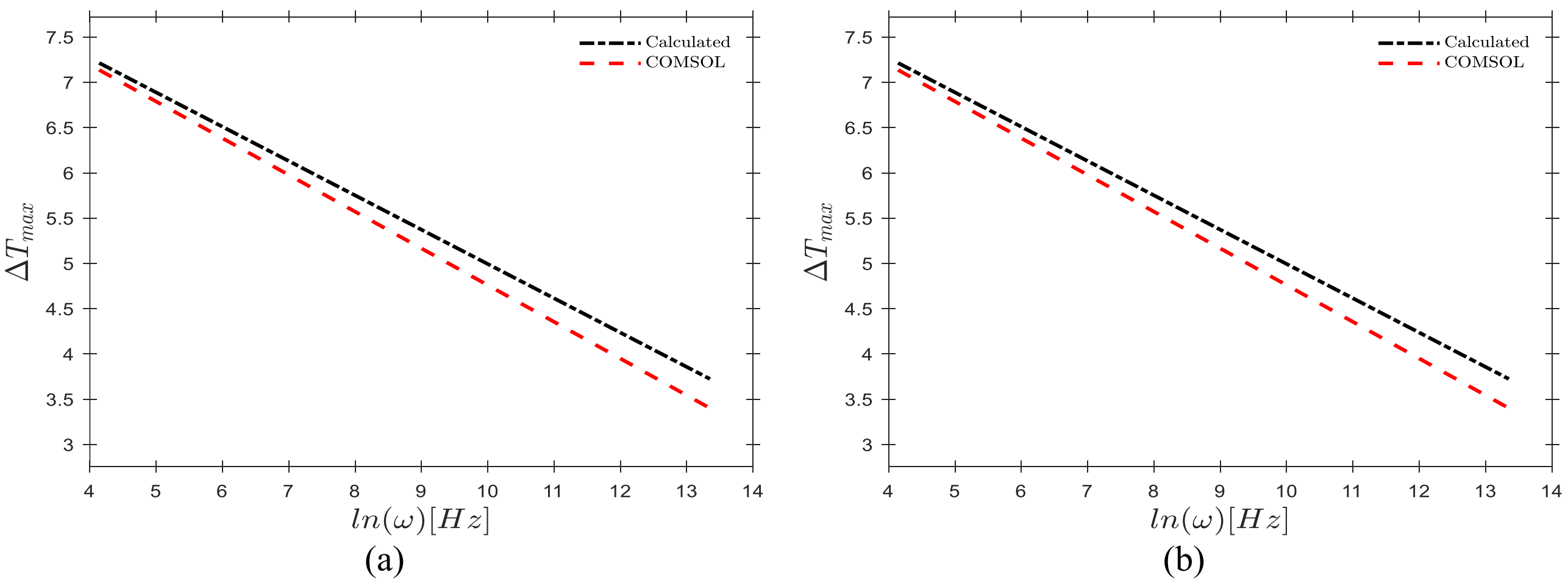

Figure 4: Case A - Linear regression curves of the obtained from solutions of the different excitation frequencies from the numerical simulation (red – dashed lines) and analytically calculations (black dotted lines) and for the (a) $V_{3\omega}$ voltages (b) peak oscillation temperature amplitudes, $\overline{|\Delta T|}$.

Based on the $\partial(V_{3\omega})/(\ln \omega)$ or the associated $\partial(|\overrightarrow{\Delta T}|)/(\ln \omega)$ gradients, the specimen's estimated thermal conductivity obtained were $\kappa_{||,calc} = 281.23\ W/m \cdot K$ and $\kappa_{||,FEA} = 284.65\ W/m \cdot K$ which were comparable to the actual specimens thermal conductivity $\kappa_{||} = 285\ W/m \cdot K$.

**2. Case B** – *Different material properties for both substrate and thin film layers*

The same procedures given for the previous case (Case A) were followed for Case B considering different film and substrate material properties to obtain the field solutions. Although the results show similar behavior in the thermal gradients and temperature oscillation trends, the temperature magnitudes are generally higher for the later scenario, i.e. case B (cf. Figure 5 (a) & (b)).

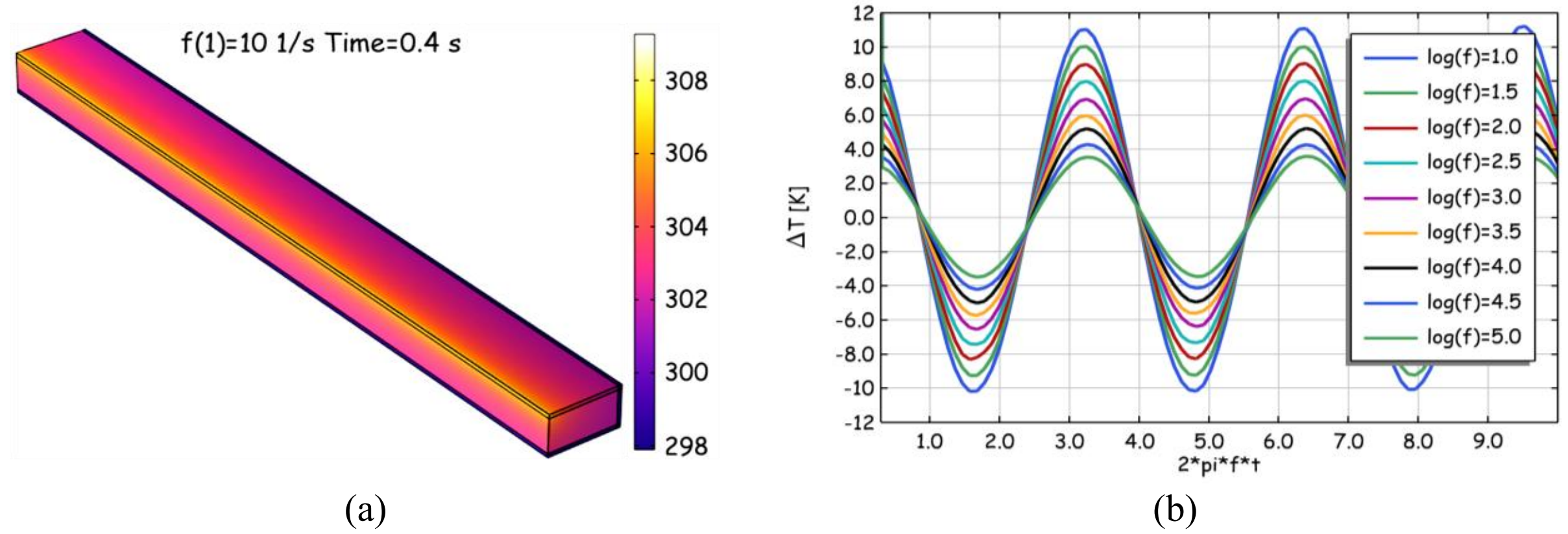

Figure 5: Case B- (a) Typical temperature distribution of the sample preparation at a frequency of $f = 10Hz$ and at time $t = 0.4s$, (b) temperature oscillations transient profiles for the different excitation frequencies with the times axis normalized with the frequency (i.e. $\omega t$).

Likewise, the numerical and analytical results of the linear regression analysis for the $V_{3\omega}$ voltages and the peak oscillation temperature amplitudes $\overline{|\Delta T|}$ are seen to be relatively higher for case B and the analytical and numerical trends are seen to converge with increased excitation frequencies (cf. Figure 6 (a) & (b)).

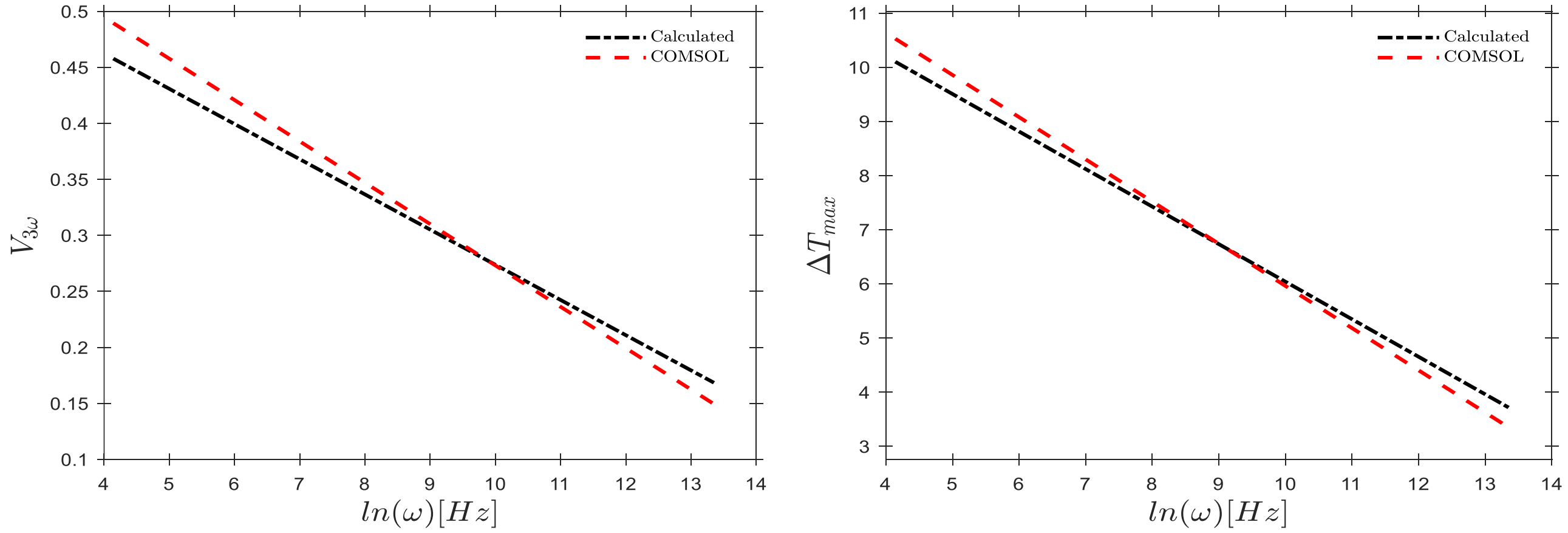

(c) (d)

Figure 6: Case B - Linear regression curves of the obtained from solutions of the different excitation frequencies from the numerical simulation (red – dashed lines) and analytically calculations (black dotted lines) and for the (a) $V_{3\omega}$ voltages (b) peak oscillation temperature amplitudes, $\overline{|\Delta T|}$.

The average specimen's thermal conductivity obtained using the gradients of the regression curves are respectively $\kappa_{||,calc} = 156.45\ W/m \cdot K$ and $\kappa_{||,FEA} = 140.00\ W/m \cdot K$ which were comparable to the actual average specimens thermal conductivity $\kappa_{||} = 155\ W/m \cdot K$, although the analytical results had aligned more closely with the actual value.

**Obtaining the Individual Layer Thermal Conductivities**

To demonstrate the validity of the NR approach in backtracking the individual layer conductivities for the three-layer system comprising the AlN-film, nanofilm and Si-substrate, three minimum Design of Experiment (DoE) instances are required. Firstly, we assume the conductivities are known a-priori to determine the $\omega - V_{3\omega} / \ \omega - \overline{|\Delta T|}$ curve as given in Figure 7 below which ordinarily would be determined from experiment.

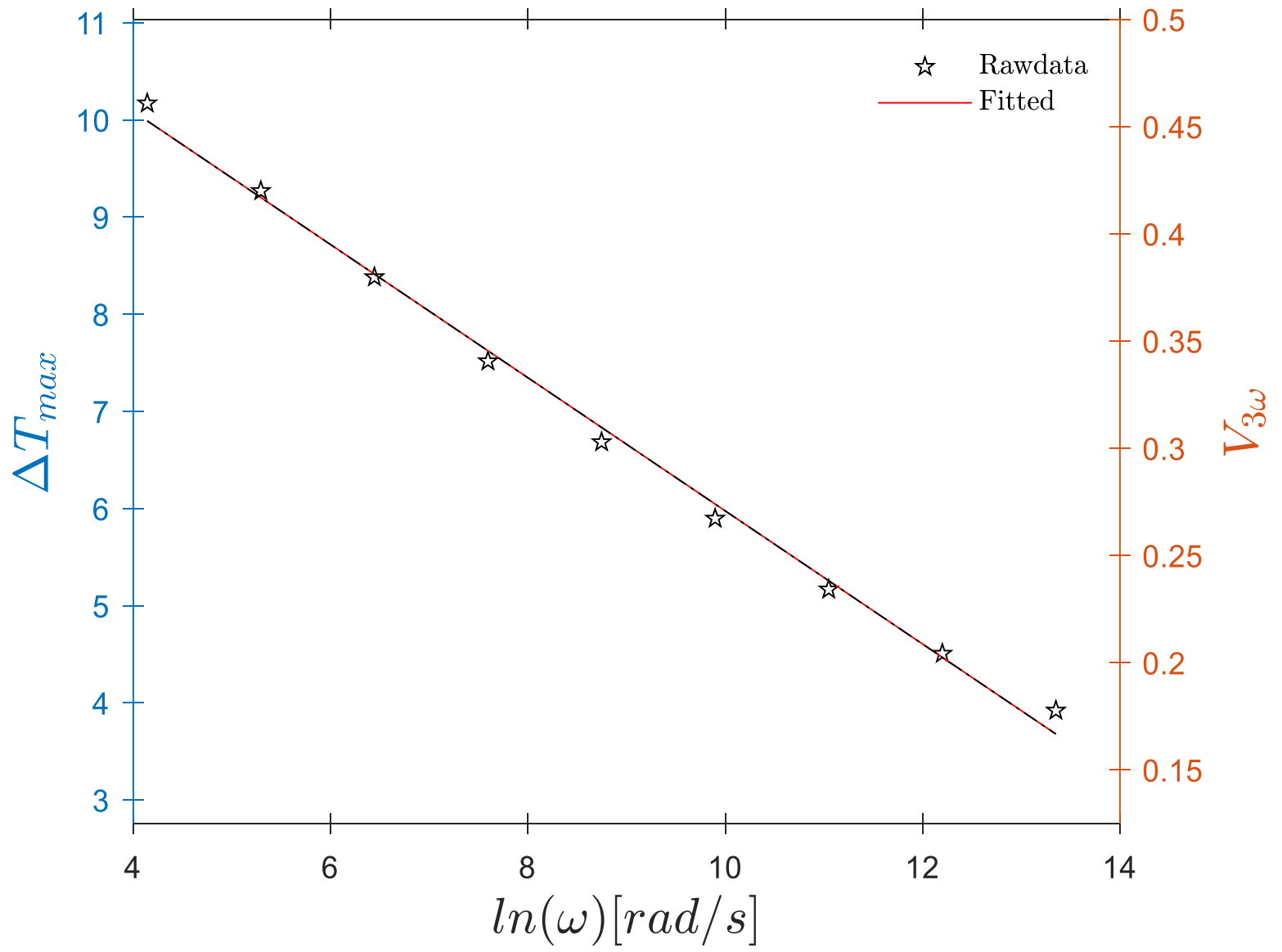

Figure 7: $\ln \omega - \overline{|\Delta T|} - V_{3\omega}$ fitted profile for the three-layer system

The three logarithmic evenly spaced DoE value sets, i.e. frequencies $f$ and corresponding $\overline{|\Delta T|}$ and $V_{3\omega}$ values and the resulting derived conductivities obtained from the NR scheme are provided in Table *4* below.

Table 4: Result of the Inverse Analysis

| $Freq.$ $[Hz]$ | $\overline{|\Delta T|}$ $[^oC]$ | $V_{3\omega}$ $[V]$ | $\kappa_{calc..}$ $[\mathrm{W/m-K}]$ | $\kappa_{act.}$ $[\mathrm{W/m-K}]$ |
|---|---|---|---|---|
| $10^1$ | 10.1710 | 0.4609 | 285.00 | 285 |
| $10^3$ | 6.6845 | 0.3029 | 119.99 | 120 |
| $10^5$ | 3.9213 | 0.1777 | 142.02 | 142 |

The result shows very good alignment between the actual layer conductivities, $\kappa_{act.}$ with values calculated from the inverse NR analysis, $\kappa_{calc.}$ The NR exhibits quadratic convergence and terminates at an error tolerance $err = \|R_j^-\| = 7.64 \times 10^{-10}$.

## Conclusion

In conclusion, the $3\omega$ method for determining the thermal conductivity of thin films has been numerically investigated in this white paper for Aluminium Nitride (AIN) thin film padded onto a Silicon (Si) substrate material. While the method is well established, the paper revisits and exposes a computational approach for validating results obtained experimentally. More importantly, the report highlights a potential method for obtaining the individual layer parallel conductivities of a multilayer system by combining a design of experiment (DoE) approach with the analytical expression of equation 10 above. In the current study, the ratio of perpendicular to parallel layer conductivities has been assumed a constant value of unity, i.e. $\kappa_{R_i} = \kappa_{\perp_i}/\kappa_{\parallel_i} = 1$. The approach can be extended to obtain the actual layer perpendicular thermal conductivities in combination with the parallel quantities utilizing the NR method by obtaining additional derivatives of the oscillation temperature amplitudes, $\overline{|\Delta T|}_i$ with the layer perpendicular conductivities $\kappa_{\perp_j}$ (see Appendix B).

## Appendix A

*Temperature Oscillation Function for Different Specimen Configurations*

For a one-dimensional line heater at the center of an infinite cylindrical specimen (cf. Figure 8 (a)) , the temperature oscillation across the film specimen is given in cylindrical coordinate as [3]

$$\Delta T_r\,(t,z) = \frac{P_0}{2\pi\kappa\, l_f} e^{i2\omega t} J_0(qr) \qquad 22$$

where $J_0$ is the zeroth order Bessel function of the second kind. For a one-dimensional line heater at the surface center of an infinite half-cylindrical specimen (cf. Figure 8 (b)) the above equation 22 is reduced by a factor of 2, i.e. $\Delta T_{r,1/2} = 2\Delta T_r$ . In this case the thermal penetration depth $d_{rt}$ is a function of the wavenumber $q$ and is given as $d_{th} = |q|^{-1}$ [3].

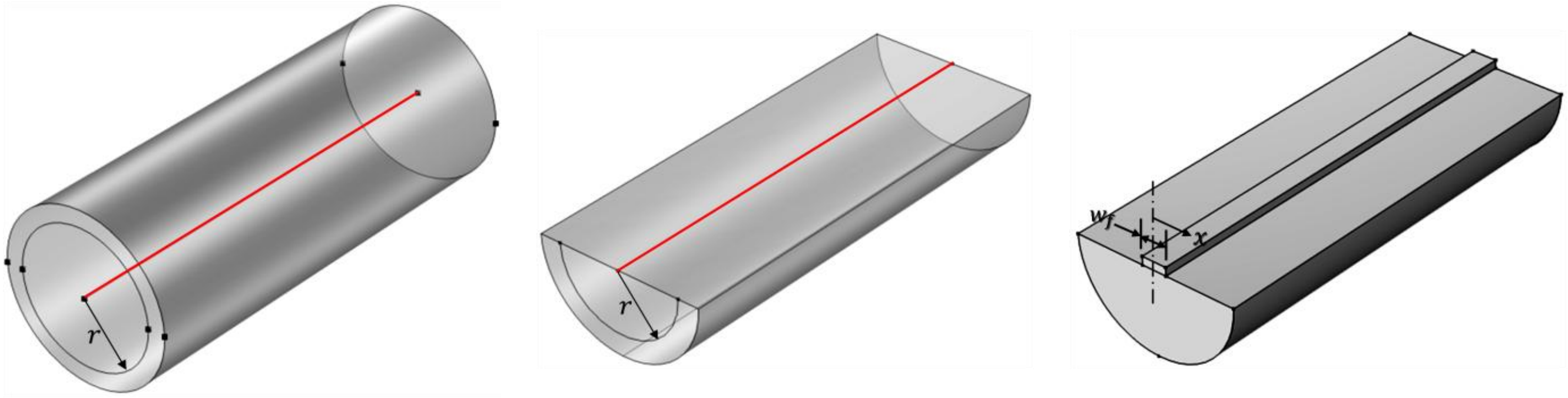

(a) (b) (c)

Figure 8: One-dimensional line heater at (a) the center of an infinite cylindrical specimen (b) the mid-surface of an infinite half -cylindrical specimen (c) finite width heater at the mid-surface of an infinite half - cylindrical specimen.

For a finite width heater at the mid-surface of an infinite half-cylinder (cf. Figure 8 (c)), the spatial oscillation temperature amplitude along the $x$ direction but at the film's surface ($z = 0$) is given as [1,3].

$$\Delta T(x) = \frac{P_0}{2\pi\kappa\, l_f}\int_0^\infty \frac{2\sin(\lambda w_f/2)\cos(\lambda x)}{\lambda w_f/2\sqrt{\lambda^2+q^2}}d\lambda \qquad 23$$

The average oscillation temperature amplitude over the width of the nanostrip is given as [3]

$$\overrightarrow{\Delta T} = \frac{1}{w_f/2}\int_0^{w_f/2} \Delta T(x)dx = \frac{P_0}{2\pi\kappa\, l_f}\int_0^\infty \frac{2\sin^2(\lambda w_f/2)}{\lambda w_f/2\sqrt{\lambda^2+q^2}}d\lambda \qquad 24$$

For $w_f|q|/2 \ll 1$, and for thin films padded on a substrate, the above expression coupled with the thermal effect of the film simplifies to [3]

$$\overrightarrow{\Delta T} \approx \overrightarrow{\Delta T_s} + \Delta T_f = -\frac{P_0}{2\pi\kappa_s l_f}[\ln(\omega)+\eta] + \frac{P_0 d_f}{w_f\kappa_f l_f}, \qquad \eta = \ln\left(\frac{w_f^2}{4\alpha}\right) + i\frac{\pi}{2} - \ln(2) \qquad 25$$

where subscript 's' and 'f' represent the substrate and film specimen respectively. Given a known measured value of $\overline{|\Delta T|}$ , equation 25 above can be used to determine the films thermal conductivity $\kappa_f$. The method is valid for ${}^{5w_f}/_{2} < |q|^{-1} < {}^{d_s}/_{5}$ or ${}^{25\alpha}/_{2d_s^2} < \omega < {}^{2\alpha}/_{25w_f^2}$

## Appendix A

*Obtaining derivatives of* $\overline{|\Delta T|}$ *w.r.t. the perpendicular layer conductivities,* $\kappa_{\perp_i}$.

To obtain the derivatives of $\overline{|\Delta T|}$ w.r.t. $\kappa_{\perp_i}$, the same set of equations given in equations 11-21 would apply however, equations 13, 18 and 19 are modified to equations 26, 27, 28 respectively as given below.

$$\frac{d}{d\kappa_{\parallel_j}}\overline{|\Delta T|}_i = -\frac{P_{rms}}{\pi l_h \kappa_{\parallel_1}}\int_0^\infty \frac{\partial}{\partial \kappa_{\perp_j}} f_i\left(\kappa_{\parallel_k}, \lambda\right) d\lambda \qquad 26$$

$$\frac{\partial B_i}{\partial \kappa_{\perp_j}} = \delta_{ij}\frac{1}{2B_j}\frac{\lambda^2}{\kappa_{\parallel_j}} \qquad 27$$

$$\frac{\partial \psi_i}{\partial \kappa_{\perp_j}} = \frac{\sinh(2\psi_i)}{2}\begin{cases} -\dfrac{1}{B_j}\dfrac{\partial B_j}{\partial \kappa_{\perp_j}} & j = i \\ \left[\dfrac{1}{A_j}\dfrac{\partial A_j}{\partial \kappa_{\perp_j}} + \dfrac{1}{B_j}\dfrac{\partial B_j}{\partial \kappa_{\perp_j}}\right] & j = i+1 \\ \dfrac{1}{A_{i+1}}\dfrac{\partial A_{i+1}}{\partial \kappa_{\perp_j}} & i+1 < j < n-1 \end{cases} \qquad 28$$